\documentclass[journal]{IEEEtran}
\usepackage{amsmath}
\usepackage{subcaption}
\usepackage{graphicx}
\usepackage{dcolumn}
\usepackage{bm}

\ifCLASSINFOpdf
\else
\fi
\begin{document}
%
% paper title
% can use linebreaks \\ within to get better formatting as desired
\title{Benchmarking  Neural Networks\\For Quantum Computations}
%
%
% author names and IEEE memberships
% note positions of commas and nonbreaking spaces ( ~ ) LaTeX will not break
% a structure at a ~ so this keeps an author's name from being broken across
% two lines.
% use \thanks{} to gain access to the first footnote area
% a separate \thanks must be used for each paragraph as LaTeX2e's \thanks
% was not built to handle multiple paragraphs
%

\author{

    \hfill

        \IEEEauthorblockN{
     Nam H. Nguyen and E.C. Behrman
    }\\
    \IEEEauthorblockA{Department of Mathematics and Physics, Wichita State University\\
    Wichita, KS, USA  67260-0033}

    \hfill

    \IEEEauthorblockN{
    Mohamed A. Moustafa and J.E. Steck
    }\\
    \IEEEauthorblockA{Department of Aerospace Engineering,Wichita State University\\
    Wichita, KS, USA  67260-0044}

    \thanks{M.A. Moustafa is now with Rescale, 33 New Montgomery St., Suite 950, San Francisco, CA 94105}% <-this % stops a space
    % \thanks{J. Doe and J. Doe are with Anonymous University.}% <-this % stops a space
    % \thanks{Manuscript received April 19, 2005; revised January 11, 2007.}

}

% note the % following the last \IEEEmembership and also \thanks -
% these prevent an unwanted space from occurring between the last author name
% and the end of the author line. i.e., if you had this:
%
% \author{....lastname \thanks{...} \thanks{...} }
%       ^------------^------------^----Do not want these spaces!
%
% a space would be appended to the last name and could cause every name on that
% line to be shifted left slightly. This is one of those "LaTeX things". For
% instance, "\textbf{A} \textbf{B}" will typeset as "A B" not "AB". To get
% "AB" then you have to do: "\textbf{A}\textbf{B}"
% \thanks is no different in this regard, so shield the last } of each \thanks
% that ends a line with a % and do not let a space in before the next \thanks.
% Spaces after \IEEEmembership other than the last one are OK (and needed) as
% you are supposed to have spaces between the names. For what it is worth,
% this is a minor point as most people would not even notice if the said evil
% space somehow managed to creep in.

% The paper headers
\markboth{Journal of \LaTeX\ Class Files,~Vol.~6, No.~1, January~2007}%
{Shell \MakeLowercase{\textit{et al.}}: Bare Demo of IEEEtran.cls for Journals}
% The only time the second header will appear is for the odd numbered pages
% after the title page when using the twoside option.
%
% *** Note that you probably will NOT want to include the author's ***
% *** name in the headers of peer review papers.     ***
% You can use \ifCLASSOPTIONpeerreview for conditional compilation here if
% you desire.

% If you want to put a publisher's ID mark on the page you can do it like
% this:
%\IEEEpubid{0000--0000/00\$00.00~\copyright~2007 IEEE}
% Remember, if you use this you must call \IEEEpubidadjcol in the second
% column for its text to clear the IEEEpubid mark.

% use for special paper notices
%\IEEEspecialpapernotice{(Invited Paper)}

% make the title area
\maketitle

\renewcommand{\arraystretch}{1.3}

\begin{abstract}

The power of quantum computers is still somewhat speculative. While they are certainly faster than classical ones at some tasks, the class of problems they can efficiently solve has not been mapped definitively onto known classical complexity theory. This means that we do not know for which calculations there will be a ``quantum advantage,'' once an algorithm is found. One way to answer the question is to find those algorithms, but finding truly quantum algorithms turns out to be very difficult. In previous work over the past three decades we have pursued the idea of using techniques of machine learning to develop algorithms for quantum computing.  Here we compare the performance of standard real- and complex-valued classical neural networks with that of one of our models for a quantum neural network, on both classical problems and on an archetypal quantum problem: the computation of an entanglement witness. The quantum network is shown to need far fewer epochs and a much smaller network to achieve comparable or better results.

\end{abstract}

% IEEEtran.cls defaults to using nonbold math in the Abstract.
% This preserves the distinction between vectors and scalars. However,
% if the journal you are submitting to favors bold math in the abstract,
% then you can use LaTeX's standard command \boldmath at the very start
% of the abstract to achieve this. Many IEEE journals frown on math
% in the abstract anyway.

% Note that keywords are not normally used for peerreview papers.

% For peer review papers, you can put extra information on the cover
% page as needed:
% \ifCLASSOPTIONpeerreview
% \begin{center} \bfseries EDICS Category: 3-BBND \end{center}
% \fi
%
% For peerreview papers, this IEEEtran command inserts a page break and
% creates the second title. It will be ignored for other modes.
\IEEEpeerreviewmaketitle

\section{Introduction}

For decades after electronic computers were invented, they exponentially increased in power and decreased in size, and the looming advent of the quantum limit was seen as a threat to progress. But in 1982 Feynman published a paper \cite{feynman1} in which he argued that quantum computing could be as much an opportunity as an obstacle. In 1994 Shor published the first quantum algorithm \cite{shor}, a systematic method to factorize integers. On a classical computer factorization is approximately exponential, that is, the time it takes is roughly exponential in the size of the integer $N$. With Shor's algorithm, running on a quantum computer, the time necessary is polynomial in the logarithm of $N$, which is  much faster for very large integers. This is an important application, because most encryption relies on the fact that while multiplication is easy and rapid, the inverse problem of factorization is not. Were a macroscopic quantum computer to be built, this would no longer be true. Shortly thereafter Lov Grover \cite{grover} found a quantum algorithm for data base searches.  Here the performance improvement is from $O(N)$ to $O(\sqrt(N))$, a quadratic speedup. Clearly at least for some problems, quantum computing can provide enormous advantages. Quantum computers are at least as fast as classical computers in general \cite{bernstein}, and widely believed to be much faster, perhaps exponentially faster, if only we can find the correct algorithms. Very recently a quantum advantage has been proved in a limited context  \cite{bravyi}. Other researchers have found quantum advantage to be ``elusive''\cite{Ronnow}. A number of researchers have proved the existence of quantum codes in a variety of contexts \cite{existence, existence2}. But existence is not construction, much less optimization. There is a need for direct comparison, and benchmarking.

Quantum Neural Networks (QNN) are an alternative approach to quantum algorithmic computing. They were first proposed independently by several researchers \cite{orig1, orig2, orig3} decades ago. More recently a number of groups have developed the idea much further. Schutzhold \cite{schutzhold} and Trugenberger \cite{trugenberger} have investigated QNNs for pattern recognition. Schuld et al. \cite{schuld}
%and Wang \cite{wang}
have looked at the related problem of linear regression.  Some reviews have recently appeared \cite{schuldrev, arunaschalam, Biamonte}. Combining quantum computing and machine learning can benefit both fields: Quantum hardware can accelerate learning \cite{Aimeur} over that for classical networks, while learning techniques can obviate the perennial problem of design of algorithms that take advantage of the quantum nature of reality  \cite{2002, 2008, chen, Hentschel}, and take advantage of bootstrapping for scaleability \cite{multi qubit, wiebe}. Some experimental small-scale demonstrations have been made, on photonic systems \cite{Cai}, and on IBM's publically available machines \cite{upcoming}.

Here, we explore the power and the complexity of a QNN, in both classical and quantum computations.  We apply standard real- and complex-valued classical neural networks to two well-understood classical tasks of computation: a single-layer perceptron, and iris identification. The complex-valued network, it will be seen, is considerably more powerful than a comparable real-valued network, at both classical tasks, while a fully quantum mechanical network does even better.

More importantly, a QNN can efficiently do fully {\it quantum} calculations.  Of course, any classical computer can simulate a quantum system, and in that sense ``do'' a quantum calculation, but with a physical quantum computer we do not need to set up a simulation: the quantum evolution of the system itself does the calculation. As noted above, we are only just beginning to get beyond simulations of QNNs. It is important to map out what kinds of calculations we will want to implement. But what are some truly quantum calculations? 

One important quantity is the ``entanglement'' of a quantum system. Entanglement is essentially quantum correlation, a correlation stronger than anything possible classically. The name derives from the fact that if two subsystems are entangled, knowledge of the one subsystem gives knowledge of the other. It is well known that the power of quantum computation relies heavily on the degree of entanglement \cite{neilsen}.  Previously, we have shown \cite{2008} that a QNN of this type can successfully calculate a general experimental entanglement witness, applicable to mixed as well as pure states, which can be bootstrapped to multiqubit systems \cite{multi qubit}, and which is robust to noise and to decoherence \cite{2 qubit noise}. We therefore compare also the performance of the classical neural nets on this problem. Our results show that a QNN is considerably better able to compute the entanglement witness than a classical neural network, with much smaller errors and much faster training, and generalized successfully from a much smaller training set.

The organization of the paper is as follows.  In Section II, we define each of the three types of neural network that we are comparing: Classical real valued, classical complex, and fully quantum mechanical. Both kinds of classical net are universal approximators. In the fully quantum case presented here, we do not yet have theorems that show precisely how the complexity and power compare; hence this present study. In Section III we compare the performance of each net on classical benchmarking problems. In Section IV we compare their performance on a purely quantum problem: characterization of the entanglement of a quantum system. In Section V we present our conclusions.

\section{Neural Network Approaches }
\subsection{Classical real-valued neural networks}

Standard algorithmic computers are ``Turing complete'', that is, can be programmed to simulate any computer algorithm. Classical, Real-Valued Neural Networks (RVNN) are also ``universal approximators.''  This means essentially that an RVNN can approximate any continuous function that maps inputs to outputs.  We will present (though not prove, since the proofs are both known and not our own) the theorems that explain explicitly what this means, in order to be able to compare with a QNN.  Proofs can be found in a number of sources  \cite{cybenko}. \\

A real-valued neural network with a continuous sigmoidal activation function $\sigma: \mathbf{R}^n \to \mathbf{R} $ can be represented as
\begin{equation}
G(x) = \sum_{j=1}^{N_{l}} \alpha_{j} \sigma \big( {\bf w}_{j}^{T} {\bf x} + \theta_{j} \big)
\end{equation}
\noindent where the sum is over the neurons in the hidden layer, $ {\bf x}, {\bf w}_{j} \in \mathbf{R}^{n} $ are the inputs and weights, respectively;  $\alpha , \theta \in \mathbf{R}$ are the scaling factors and biases of the activation function, and where  $ {\bf w}_{j}^{T} $ represents the transpose of ${\bf w}_{j}$.  \\

It should be noted that  $\sigma$ is discriminatory. This means for a measure $\mu \in M(I_{n}) $
\begin{equation}
\int_{I_{n}} \sigma \big( {\bf w}^{T} {\bf x} + \theta \big) d\mu(x)  = 0
\end{equation}
\noindent for all ${\bf x} \in \mathbf{R}^{n}$ and $\theta \in \mathbf{R}$ implies that $\mu =0 $. Here $M(I_{n})$ is the space of finite, signed, regular Borel measures on the unit interval, $I_{n}$. This requirement is to ensure that the activation cannot send the affine space ${\bf w}^{T}{\bf x} + \theta$ to the set of measure zero. \\

\textbf{Theorem 1:} Let $\sigma$ be any continuous sigmoidal function. Then finite sums of the form
\begin{equation}
G(x) = \sum_{j=1}^{N_{l}} \alpha_{j} \sigma \big( {\bf y}_{j}^{T} {\bf x} + \theta_{j} \big)
\end{equation}
are dense in the continuous functions on the unit interval, $C(I_{n})$ \\

Theorem 1 tells us that as long as the mapping between the inputs and targets can be described by a continuous function, $G: \mathbf{R}^{n} \to \mathbf{R} $, then for a real-valued neural network with just one hidden layer and any continuous sigmoidal function as the activation function there exist weights $w$, bias $\theta$ and scalar $\alpha$ to approximate this continuous map to any desired accuracy. When the activation is a bounded, measurable sigmoidal function then \cite{cybenko} the finite sum $G(x) = \sum_{j=1}^{N_{l}} \alpha_{j} \sigma \big( {\bf y}_{j}^{T} {\bf x} + \theta_{j} \big)$ is dense in $L^{1}(I_{n})$. \\

An (in house) implementation of a real-valued artificial neural network was used as a baseline. The software NeuralWorks Professional II \cite{NeuralWorks} was also used to verify that the results obtained from commercially available software were comparable to our own implementation.

    % \begin{figure}
    %     \centering
    %     \includegraphics[width=0.4\textwidth]{mapping_figure.eps}
    %     \caption{Range }
    %     \label{fig:mapping}
    % \end{figure}

% \vspace{1 cm}

\subsection{Complex-valued neural networks}

We now extend from $ \mathbf{R}^{n}$ to $\mathbf{C}^{n}$. Clearly it ought to be true that if a calculation can be done in $\mathbf{R}^n$ it can be done in $\mathbf{C}^n$, since the real valued neural network is a special case of the Complex-Valued Neural Net (CVNN). And it is true that the complex-valued net is a universal approximator also, though we need to be a little careful.

\textbf{Theorem 2:} Let $\sigma: \mathbf{C}^n \to \mathbf{C} $ be any complex continuous discriminatory function. Then the finite sums of the product of the form
\begin{equation}
F(z) = \sum_{k=1}^m \beta_k \prod_{l=1}^{s_k} \sigma \big( {\bf W}_{kl}^T {\bf z} + \theta_k \big)
\end{equation}
are dense in $C(I_n)$. Here $\beta_k, \theta_k \in \mathbf{C}$, $ W_{kl}, {\bf z} \in \mathbf{C}^n $ , $m$ is the number of neurons in the hidden layer, and $s_k$ is an arbitrary number depending on how many product terms one wants to do. The product is needed to ensure that the set $ \tau = \big\{ F: \mathbf{C}^n \to \mathbf{C} \big\} $ is an algebra since it will be closed under multiplication \cite{cnn}. This is the difference between Theorem 1 and Theorem 2. Similarly to the real-valued networks, if the activation function, $\sigma: \mathbf{C}^n \to \mathbf{C}  $ is any bounded measurable discriminatory function then the finite sum $F(z)=\sum_{k=1}^m \beta_k \prod_{l=1}^{s_k} \sigma \big( {\bf W}_{kl}^T {\bf z} + \theta_k \big) $ is dense in the Lebesgue integrable functions on the unit interval, $L^1(I_n)$. \\

In the complex valued case, the activation function may have essential singularities. But if  $\sigma: \mathbf{C}^n \to \mathbf{C}  $ is any complex function having an isolated and essential singularity then the finite sum above will be dense in compact subsets of analytic deleted neighborhood of the singularity. \\

This shows that there exists a representation of the CVNN  that can solve essentially any problem the RVNN can do. Indeed, we expect that the CVNN will be more powerful or at least more efficient. The archetypal example is the XOR. Since it is a nonlinear problem, a real-valued net needs a hidden layer to solve, but the complex-valued net can do this in a single layer \cite{complex network text}.  \\

We are currently working on a new QNN architecture, to which we will be able to apply this theorem directly, that is, a QNN that will be a universal approximator \cite{nams frat}. Here, we use instead simpler versions for the initial benchmarking work, both of the QNN and of the CVNN, which have already been explored in the literature.  The implementation of the CVNN used here is largely based on the work of Aizenberg \cite{complex network text}. The major difference from the RVNN is that  the neuron inputs, outputs, weights, and biases are numbers in the complex plane. In the RVNN signals are only scaled by their respective weights, while the CVNN allows the scaling and rotation of the signals. This leads to higher functionality in more compact architectures. Similar to a conventional RVNN, the input signals to any given neuron are summed by their respective weights:

    \begin{equation}
    z = \sum_n {\bf w}_n \cdot {\bf x}_n
    \end{equation}

\noindent where ${\bf w}$ and ${\bf x}$ are the complex-valued weights and signals. An activation function is then applied to the sum $z$ at the neuron. The complex activation function $P$ takes the inputs and maps them to a unit circle on the complex plane which constitutes the signal from the neuron to the consequent layer.

    \begin{equation}
    P(z) = e^{i \cdot arg(z)}
    \end{equation}

%\noindent where as usual $i$ refers to the imaginary unit.
\noindent The network's scalar inputs and outputs/targets are condensed to the range [0,1] and mapped as points on the complex unit circle using the following mapping function:

    \begin{equation}
    M(r) = e^{i \cdot \pi r}
    \end{equation}

Unlike the RVNN, with this architecture there is no need to take the derivative of the activation function or to apply gradient descent in order for this network to learn. The resulting error from each training pair can be represented as the vector from the output to the target in the complex plane. In other words, given the target $t$ and the output $z$ of an output neuron, the error can be represented as the vector $e$

    \begin{equation}
    e = t - z = \sum_n \Delta {\bf w}_n \cdot {\bf x}_n
    \end{equation}

\noindent given that the resulting error is a contribution of all of the signals to the output neuron. Here $\Delta w_n$ is the contribution of each individual weight to the error. Dividing the error equally among the $N$ incoming weights we arrive at the following required weight adjustments for each training pair per epoch.

    \begin{equation}
    \Delta {\bf w}_n = \frac{e}{N} \cdot {\bf x}_n^{-1}
    \end{equation}

For each subsequent layer the error is backpropagated by dividing the error across the weights relative to the values of the signals they carry. Thus the error gets diluted as it propagates backwards, yet this is applicably manageable given that the higher functionality of the CVNN favors shallower networks for conventional problems.

\subsection{Quantum Neural Network}

Because the quantum benchmark calculation chosen here is that of an entanglement witness, we consider the model for a QNN we first proposed in 2002 for this purpose \cite{2002, 2008}. The system is prepared in an initial state, allowed to evolve for a fixed time, then a (predetermined) measure is applied at the final time. The adjustable parameters necessary are the (assumed time-dependent) couplings and tunneling amplitudes of the qubits; the discriminatory function is provided by the measurement, which we may take to be any nontrivial projection convenient without loss of generality. For the two-qubit system, we choose as output the square of the qubit-qubit correlation function at the final time, which vaires between zero (uncorrelated) and one (fully correlated.) Thus we are considering the map
\begin{equation}
f: C^n \rightarrow [0,1]
\end{equation}
This structure lends itself immediately to the perceptron problem, as well as the entanglement problem.

The network input is the initial state of the system, as given by the density matrix $\rho(0)$. The signal is then propagated through time using Schrodinger's equation
\begin{equation}\label{Schr gen}
\frac{d\rho}{dt}= \frac{1}{i \hbar}[H,\rho]
\end{equation}
where $H$ is the Hamiltonian. Note that we work with the density matrix representation because it is computationally easier than using the state vector (or wave vector) representation. The density matrix form is more general, allowing mixed as well as pure states, but here we will in fact use only pure states. In terms of the state vector of the system $|\psi\rangle$, the density matrix is given by $\rho = |\psi\rangle \langle \psi|$.

For a two-qubit system, we take the Hamiltonian to be:
\begin{equation}
H= \sum_{\alpha=1 }^2 K_{\alpha}X_{\alpha} + \epsilon_{\alpha}Z_{\alpha} + \sum_{\alpha \neq \beta =1 }^2 \zeta_{\alpha\beta}Z_{\alpha} Z_{\beta}
\end{equation}
where $\{X,Z\}$ are the Pauli operators corresponding to each qubit, $\{K\}$ are the amplitudes for each qubit to tunnel between states, $\{\epsilon\}$ are the biases, and $\{\zeta\}$ are the qubit-qubit couplings. Generalization to a fully connected $N$ qubit system is obvious. We choose the usual charge basis, in which each qubit's state is $\pm$1, corresponding to the Dirac notation $|0\rangle$ and $|1\rangle$, respectively.

By introducing the Liouville operator, $L= \frac{1}{\hbar} \big[\cdots, H\big]$, Eq. (\ref{Schr gen}) can be rewritten \cite{peres} as
\begin{equation}\label{schro mod}
\frac{\partial \rho}{\partial t} = -iL\rho
\end{equation}
which has the general solution of
\begin{equation}\label{schro sol}
\rho(t) = e^{-iLt}\rho(0)
\end{equation}

We propagate our initial state, $\rho(0)$,  forward in time to the state $\rho(t_f)$ where we will perform a measurement on $\rho(t_f)$, which mathematically is an operator followed by a trace. That measurement, or some convenient mathematical function of that measurement, is the output of the net. That is, the forward operation of the net can be pictured as:\\
\begin{equation}\label{rhoevolmap}
\rho(0) \xrightarrow{ \frac{d\rho}{dt}= \frac{1}{i \hbar}[H,\rho] } e^{-iLt}\rho(0) = \rho(t_f) \rightarrow \mathrm{Trace} (\rho(t_f)M)=\langle M \rangle
\end{equation}
\noindent where $M$ represents the particular measurement chosen. To train the net, the parameters $K$, $\epsilon$ and $\zeta$  are adjusted based on the error, using a quantum analogue \cite{2008} of Werbos's well-known back-propagation technique \cite{werbos}.

There is, clearly, no direct correspondence between these equations for the QNN with those of either the real- or complex-valued neural network, above. The nonlinearity necessary to do learning comes entirely from the (necessarily non-linear) measurement step. We do not know exactly how the complexity of this model compares; work on these theorems is on-going \cite{namthesis}.

However, as a first step, we can investigate the question experimentally, in what follows.

\section{Benchmarking on classical problems}

We first compare the performance of our neural networks on two classical problems: logic gates and pattern classification.

%, and on a purely quantum calculation, an entanglement witness.

    \subsection{Logic Gates}

\begin{figure}
        %\centering
        \includegraphics[width=0.4\textwidth]{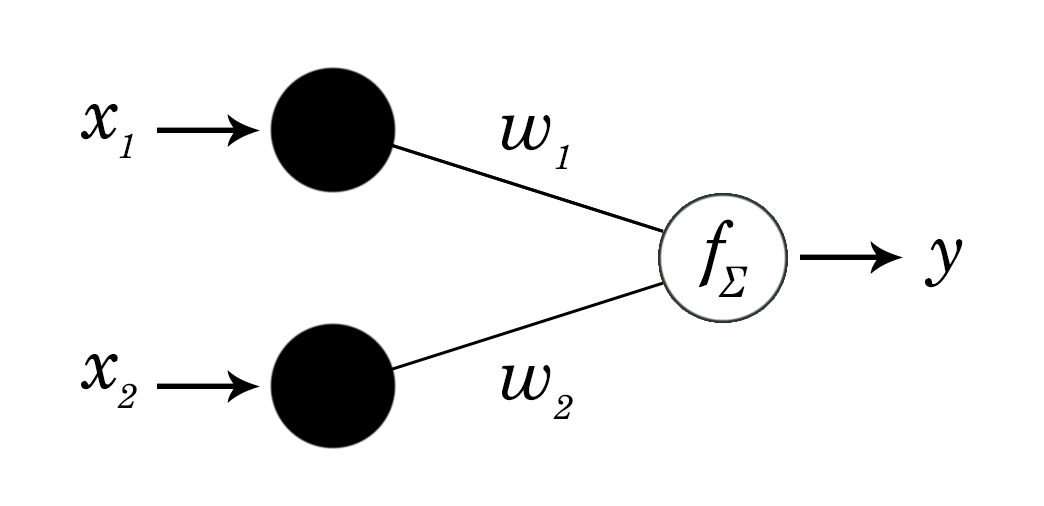}
        \caption{Single-layer perceptron architecture.}
        \label{fig:perceptron}
    \end{figure}

Perhaps the most basic of neural network architectures is the single-layer perceptron. See Fig. \ref{fig:perceptron}. The perceptron is a single output neuron connected to the input neurons by their respective weighted synapses. The real-valued perceptron is a linear classifier and therefore can approximate linearly separable logic functions such as AND and OR gates; however, because it essentially creates a single linear separator, it necessarily fails to approximate non-linear gates, like the XOR or XNOR.
%This point is made graphically in Fig.\ref{fig:separability}: a straight line can separate points for which different outputs are desired.
In contrast, a complex perceptron can calculate the XOR or XNOR, since it can create a nonlinear separation - or, equivalently, uses interference.

%    \begin{figure}[htb]
%    \centering
%      \begin{subfigure}{0.49\linewidth}
%        \centering
%        \includegraphics[width=.99\textwidth]{OR_separability.png}
%        \caption{OR}
%      \end{subfigure}%
%      \begin{subfigure}{0.49\linewidth}
%        \centering
%        \includegraphics[width=.99\textwidth]{XOR_separability.png}
%        \caption{XOR}
%      \end{subfigure}%
%      \caption{Visual representation of the linear separability of the OR gate versus the XOR gate. The dotted line represents linear classifier.}\label{fig:separability}
%    \end{figure}

\begin{figure}[htb]
    \centering
      \begin{subfigure}[b]{.49\linewidth}
        \centering
        \includegraphics[width=.99\textwidth]{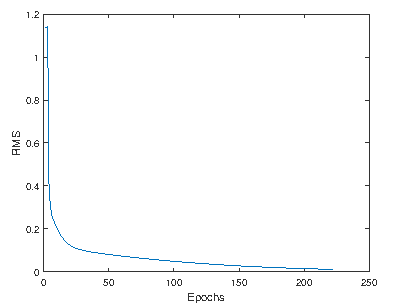}
        \caption{AND}
      \end{subfigure}%
      \begin{subfigure}[b]{.49\linewidth}
        \centering
        \includegraphics[width=.99\textwidth]{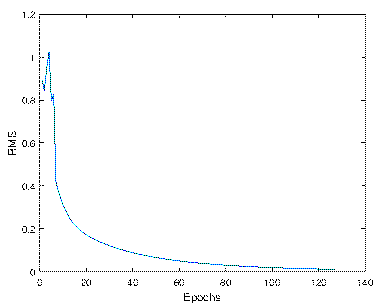}
        \caption{NAND}
      \end{subfigure}\\%
      \begin{subfigure}[b]{.49\linewidth}
        \centering
        \includegraphics[width=.99\textwidth]{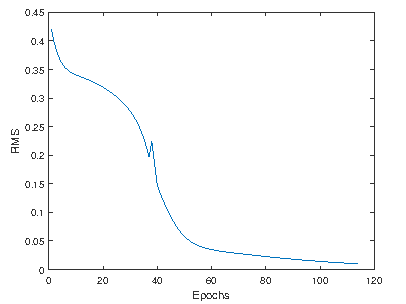}
        \caption{OR}
      \end{subfigure}%
      \begin{subfigure}[b]{.49\linewidth}
        \centering
        \includegraphics[width=.99\textwidth]{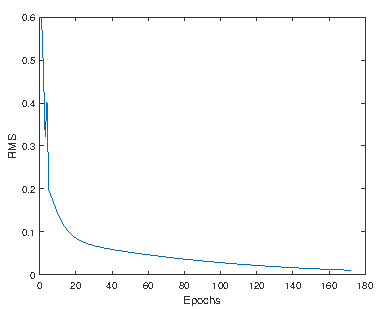}
        \caption{NOR}
      \end{subfigure}\\%
      \begin{subfigure}[b]{.49\linewidth}
        \centering
        \includegraphics[width=.99\textwidth]{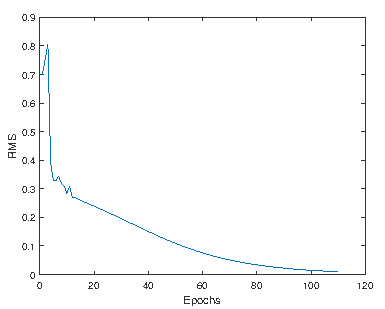}
        \caption{XOR}
      \end{subfigure}%
      \begin{subfigure}[b]{.49\linewidth}
        \centering
        \includegraphics[width=.99\textwidth]{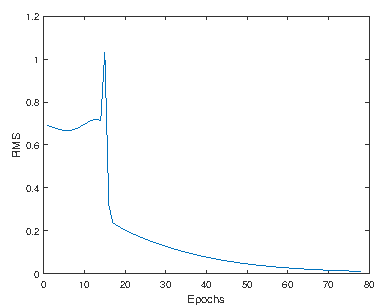}
        \caption{XNOR}
      \end{subfigure}%
      \caption{ Example of training RMS as a function of epoch for various logic gates using the  (single layer) CVNN perceptron. The RVNN and NeuralWorks networks trained similarly. }\label{fig:rms}
    \end{figure}

The single layer perceptron was implemented on all four networks. The networks were trained on the 4 possible input pairs, \{(00), (01), (10), and (11)\}, for the outputs of the AND, NAND, OR, NOR, XOR and XNOR gates until the root mean square error (RMS) reached 1\%. See Fig. \ref{fig:rms} for an example. The learning rate of each network was increased to optimal, i.e, the training speed was maximized subject to the constraint that convergence was still achieved. This was done to draw a more equitable comparison on learning speed between the RVNN and CVNN, which have different learning schemes.

To implement logic gates on our QNN model, we must specify the encoding of inputs as a function of the prepared state of the quantum system,  and outputs as a function of some measurement on the final state of the quantum system. We take the inputs as being the basis states of the two-qubit system in the so-called ``charge'' basis, that is, the states $|00 \rangle$,  $|01 \rangle$,  $|10 \rangle$, $|11 \rangle$.  We choose the output measure as being the square of the qubit-qubit correlation function at the final time, i.e, $\langle Z_{A}(t_{f})Z_{B}(t_{f})\rangle^{2}$ . An output of one would mean that the two qubits are perfectly correlated to each other at the final time; an output of zero, that they are perfectly uncorrelated. The target values for these outputs will, of course, be different for each logic gate considered. Thus, for example,  for the XNOR gate we would want an inital state of $|00\rangle$ or an inital state of $|11\rangle$ to train to a final correlated state but initial states of $|01\rangle$ or of $|10\rangle$ to train to a final uncorrelated state.  Because of the inherent continuity of the QNN any initial state close to one of the input states will necessarily produce an final state close to the corresponding output state, and the computation is robust to both noise and decoherence \cite{2 qubit noise}.
%
%     \begin{table}[h]
%        \centering
%        \label{gate}
%        \begin{tabular}{l|l|}
%%%%%        \cline{2-7}
%        \hline
%        \multicolumn{1}{|c|}{Inputs} & \multicolumn{1}{c|}{Outputs}  \\ \hline
%        \multicolumn{1}{|l|}{ $|00 \rangle$  }  &  1    \\ \hline
%        \multicolumn{1}{|l|}{ $|01 \rangle$  }  &  0  \\ \hline
%        \multicolumn{1}{|l|}{ $|10 \rangle$  }  &  0      \\ \hline
%        \multicolumn{1}{|l|}{ $|11 \rangle$  }  &   1  \\ \hline
%        \end{tabular}
%        \caption{Inputs and Outputs for the QNN model of the XNOR gate. The input is the initial, prepared, state of the two-qubit system, at $t=0$; the output, the square of the measured value of the qubit-qubit correlation function at the final time.}
%        \label{table:gate}
%    \end{table}

Table \ref{table:gates_epochs} shows the number of training epochs required by each ``optimal'' network to reach an RMS error of 1\%, where an epoch is defined as one pass through the 4 training pairs.   Note that the nonlinear gates (XOR and XNOR) cannot be done by either of the real classical nets (RVNN and NeuralWorks) with only a single layer. The QNN reached $\leq$1\% error in a single epoch.

            \begin{table}[h!]
            \centering

            \begin{tabular}{ |l|l|l|l|l| }
             \hline
             \multicolumn{5}{|c|}{No. of Epochs to RMS error $\leq$1\%} \\
             \hline
             Gate & N.Works & RVNN & CVNN & QNN\\
             \hline
             AND   & 11000 & 11146 & 222 & 1 \\
             \hline
             NAND  & 11000 & 11145 & 127 & 1 \\
             \hline
             OR    & 6000 & 5672 & 114 & 1 \\
             \hline
             NOR   & 6000 & 5671 & 172 & 1 \\
             \hline
             XOR   & n/a & n/a & 110 & 1\\
             \hline
             XNOR  & n/a & n/a & 78 & 1\\
             \hline
            \end{tabular}

            \caption{Number of epochs needed for gate training to reach RMS error $\leq$1\% for a (single layer) perceptron using RVNN, CVNN and NeuralWorks (to nearest 1000 epochs) implementations. }
            \label{table:gates_epochs}
        \end{table}

The real-valued network would not train to an RMS error below 50\% for the XOR and XNOR gates, given a number of epochs up to the order of 10\textsuperscript{6}. In addition, for the linearly separable gates the RVNN required 30-50 times more learning iterations than the CVNN to reach an RMS error of 1\%, making the CVNN training runs computationally much more efficient than RVNN. Note that the single-layer complex-valued perceptron can learn to solve the non-linearly separable XOR and XNOR gates, and do so with an efficiency at least  comparable to that for the linearly separable gates as shown in Table \ref{table:gates_epochs}, as mentioned above.

``Quantum-inspired'' networks \cite{bhattacharyya} generally are some version of a CVNN, and, therefore, do have increased power over a RVNN due to their ability to use interference: that is, their performance would be expected to be comparable to our CVNN. However a fully quantum network can do even better. Results of the training of the QNN for the linear and non-linear logic gates are shown in the last column of Table \ref{table:gates_epochs}.  Note that the QNN requires only a single epoch to learn any of the linear or the nonlinear gates, and it should be noted that the error continued to decrease below that level. This is an experimental realization of our 2015 theoretical result deriving weights for a quantum perceptron \cite{perceptron}.

    % \begin{figure}[htb]
    % \centering
    %   \begin{subfigure}[b]{0.9\linewidth}
    %     \centeringf
    %     \includegraphics[width=.99\textwidth]{XOR_targets_before}
    %     % \caption{AND}\label{fig:1a}
    %   \end{subfigure}\\%
    %   \begin{subfigure}[b]{0.9\linewidth}
    %     \centering
    %     \includegraphics[width=.99\textwidth]{XOR_targets_after} In gen
    %     % \caption{NAND}
    %   \end{subfigure}%
    %   \caption{Mapped training targets and CVNN outputs for XOR gate before and after training completion}\label{fig:1}
    % \end{figure}

    \subsection{Iris Classification}

\begin{figure}
        \centering
        \includegraphics[width=0.5\textwidth]{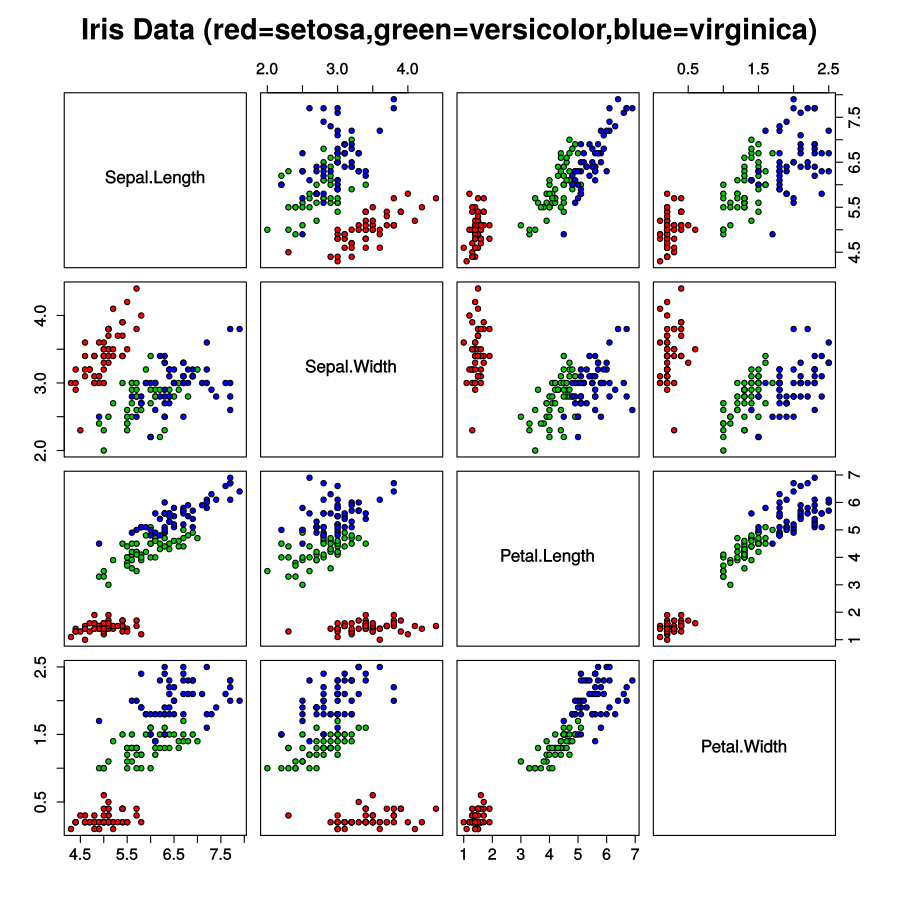}
        \caption{Iris dataset cross-input classification scatterplot  \cite{irisdataset}. }
        \label{fig:iris_scatterplot}
    \end{figure}

The other archetypal machine learning task we consider is that of pattern classification, for which one of the most common benchmarking problems is the Iris flower classification dataset \cite{irisdataset}. The multivariate dataset contains four features, as inputs: sepal length, sepal width, petal length, and petal width. The corresponding taxonomy of Iris flower (setosa, virginica or versicolor) is represented as a ``one-hot'' vector \cite{harris}. The set contains a total of 150 pairs, 50 of each species, as identified by Fisher in 1936 \cite{fisher}. The difficulty in Iris classification is well demonstrated by the scatterplot shown in Fig. \ref{fig:iris_scatterplot}: two of the the three species cluster together and require a highly non-linear separability function.

The RVNN and CVNN implementations were trained on the dataset to compare their performance for a non-linear multivariate classification problem. The dataset was divided into two randomly selected sets of 75 pairs containing an equal number of each species for training and testing. The networks were first trained on the entirety of the training set. The training pairs were then reduced to 30 and 12 pairs while keeping the testing set at 75 pairs. Both the RVNN and CVNN had a single hidden layer, that is, an architecture (4, $N_h$, 3), where the number of neurons in the hidden layer, $N_h$, was increased from three on up. Unlike the RVNN, the CVNN's testing performance improved substantially as $N_h$ was increased, up to about 100; this is the (optimized) architecture for the results reported in Table \ref{table:iris_results}. ``Testing'' accuracy in Table \ref{table:iris_results}  means classification in the correct category if the appropriate output is above 0.5; below, it is counted as incorrect. 

For implementation on the QNN, we used the feature data as the coefficients of the input state $|\psi(0)\rangle$ in the Schmidt decomposition. That is, the initial quantum state vector for each input was given by
\begin{equation}
\label{eq:purestate}
|\psi(0)\rangle = a |00\rangle + b e^{i\theta_1)} |01\rangle + c e^{i \theta_2}|10\rangle + d e^{i \theta_3} |11\rangle
\end{equation}
We set all the phase offsets $\{ \theta_{i} \}$ equal to zero (i.e., only real inputs here), and take each amplitude to correspond to a features (i.e., $a,b,c,d $ correspond to sepal length, sepal width, petal length, and petal width.) For the output, we again choose the final time correlation function, taken to be the polynomial $\frac{0.01(a^{2}+b^{2})+c^{2} + 1.8d^{2}}{a^2 + b^2 +c^2 + d^2 }$. This polynomial sorts out the three categories of flowers almost completely. It is  possible to choose a more complicated function or higher degree polynomial to create more separation but we chose the simplest polynomial for the ease of training. 
 %Most of the separation can be accomplished with only the petal data; with this function, almost complete separation is achieved. 
 
     \begin{table*}[h]
    \centering
    \begin{tabular}{l|l|l|l|l|l|l||l|l|l|}
    \cline{2-7}
    \multicolumn{1}{c|}{}  & \multicolumn{3}{c|}{Training RMS (\%)} & \multicolumn{3}{c|}{Testing RMS (\%)}  & \multicolumn{3}{c|}{Testing Accuracy (\%)}           \\ \hline
    \multicolumn{1}{|c|}{Training Pairs} & \multicolumn{1}{c|}{RVNN} & \multicolumn{1}{c|}{CVNN} & \multicolumn{1}{c|}{QNN} & \multicolumn{1}{c|}{RVNN} & \multicolumn{1}{c|}{CVNN}  & \multicolumn{1}{c|}{QNN} & \multicolumn{1}{c|}{RVNN} & \multicolumn{1}{c|}{CVNN}  & \multicolumn{1}{c|}{QNN} \\ \hline
    \multicolumn{1}{|l|}{75}&3.45  &2.06 &0.96  &3.71 &7.78 & 2.31 &100 &100 & 97.3 \\ \hline
    \multicolumn{1}{|l|}{30}&0.51  &0.41 &1.1  &4.97 &9.47 & 9.78 &93.3 &96.0 & 97.5 \\ \hline
    \multicolumn{1}{|l|}{12}&0.69  &0.09 &0.62  &11.9 &16.4 & 7.48&85.3  &94.7 & 85.5 \\ \hline
    \end{tabular}
    \caption{Iris dataset training and testing average percentage RMS and identification accuracy for different training set sizes using similar RVNN and CVNN networks, and compared with the QNN. The RVNN was run for 50,000 epochs; the CVNN, for 1000 epochs; and the QNN, for 100 epochs. See text for details of the architectures used.}
    \label{table:iris_results}
    \end{table*}

Results are shown in Table \ref{table:iris_results}. On this classification problem the advantage of the CVNN over the RVNN is not nearly as pronounced. The RMS training errors for the CVNN are consistently below those of the RVNN, but the testing errors are not. In terms of the classification the CVNN does do better, but not as significantly as with the perceptron. The reason the classification improves while the RMS error does not is that the CVNN tends to a flatter output: thresholding gives the ``correct'' answer, but the separation between classification of ``yes'' (i.e., a value near 1) and ``no'' (i.e. a value near zero) is smaller with the CVNN.

Performance in training, testing, and classification by the QNN is also comparable, to both nets. The major difference among the three is in the amount of training necessary to achieve the results. The RVNN required 50,000 epochs, while the CVNN required only 1000, and the QNN only 100. The ratios of iterations necessary is remarkably similar to the results for the perceptron, shown in Table \ref{table:gates_epochs}. In this sense we can say that the QNN is significantly more efficient at training, in classification as well as in computation.

\section{Quantum computation problem}

We now consider our networks' performance on a purely quantum calculation: an entanglement witness.

\subsection{Benchmarking with entanglement}
An outstanding problem in quantum information is quantifying one of the key elements responsible for the powerful capabilities of quantum computational systems: entanglement. Entanglement is the feature whereby two or more subsystems' quantum states cannot be described independently of each other, that is, that the state of a system AB cannot be correctly represented as the (state of A)$\otimes$(state of B), where $\otimes$ indicates the tensor product. Non-multiplicativity, or entanglement, of states, is fundamental to quantum mechanics. Indeed, all of quantum mechanics can be derived from this characteristic \cite{feynman}; in that sense, entanglement is equivalent to non-commutativity of operators, or to the Heisenberg Uncertainty Principle.

Even though quantifying entanglement has been a prominent subject of research, there is no accepted closed-form representation that is generalizable to quantum states of arbitrary size or properties. Because a quantum state is unentangled if it is separable into its component states, entanglement measures are used to determine the ``closeness'' of a quantum state to the subspace of separable product states. Most of the developed measures involve a subset of possible compositions or require extremizations which are difficult to implement \cite{optimizations}, especially for large systems. One well-known measure is the entanglement of formation which quantifies the resources utilized to construct a specific quantum state. For a two-qubit system we have an explicit formula for the entanglement of formation \cite{wootters}, but generalizations to larger systems are difficult.

The lack of a general analytic solution combined with a base of knowledge on specific sets of states suggests an opportunity for a machine learning approach.  Given that the entanglement is an inherent property of the quantum mechanical system, that system in a sense ``knows'' what its entanglement is. Thus, if we are clever, we can devise a measurement or measurements that will extract this information, and, it is hoped, somewhat more efficiently than complete determination of the quantum state (``quantum tomography'') would require, as the number of measurements goes like $2^{2n}$, where $n$ is the number of qubits. It is a theorem \cite{2008} that a single measurement cannot determine the entanglement completely even for a two qubit system; however, one measurement can serve as a ``witness'': that is, a measurement giving a lower bound to the correct entanglement.

Using a training set of only four pure states, we successfully showed \cite{2008} that this model of a QNN can ``learn'' a very general entanglement witness, which tracks very well with the entanglement of formation \cite{wootters}, and which not only works on mixed as well as pure states but can be bootstrapped to larger systems \cite{multi qubit} and, most importantly, is robust to noise and to decoherence \cite{2 qubit noise,robustness}.

Any computation done by a quantum system can be simulated; any function can be approximated. But how complex is this calculation? Theorems 1 and 2 tell us that the weights exist, but how many do we need, and are they easy to find? For the two-qubit system, this is fairly easy to characterize. Note that for the time-propagation part of the computation, each step is linear. Thus, if we have a two qubit system Eq. \ref{rhoevolmap} can be written out as follows. Let $M$ be the 4x4 matrix $M=\{m_{ij}\}$, let the time propagator $T = e^{-iLt}$ be $T=\{T_{ij}\}$, and let $\rho(0) = \{\rho_{ij}\}$.

Then $\rho(t_f) = T\rho(0) =$
   \begin{equation*}
 \begin{bmatrix}
    \sum_{j=1}^4 T_{1j}\rho_{j1}       & \sum_{j=1}^4 T_{1j}\rho_{j2} & \sum_{j=1}^4 T_{1j}\rho_{j3}  & \sum_{j=1}^4 T_{1j}\rho_{j4} \\
    \sum_{j=1}^4 T_{2j}\rho_{j1}       & \sum_{j=1}^4 T_{2j}\rho_{j2} & \sum_{j=1}^4 T_{2j}\rho_{j3}  & \sum_{j=1}^4 T_{2j}\rho_{j4} \\
    \sum_{j=1}^4 T_{3j}\rho_{j1}       & \sum_{j=1}^4 T_{3j}\rho_{j2} & \sum_{j=1}^4 T_{1j}\rho_{j3} & \sum_{j=1}^4 T_{3j}\rho_{j4} \\
    \sum_{j=1}^4 T_{4j}\rho_{j1}       & \sum_{j=1}^4 T_{4j}\rho_{j2} & \sum_{j=1}^4 T_{4j}\rho_{j3}   & \sum_{j=1}^4 T_{4j}\rho_{j4} \\
\end{bmatrix} \cdot
\end{equation*}
Therefore,
\begin{equation}\label{quadratic}
\big( Tr[M\rho(t_f)] \big)^2 = \bigg( \sum_{l=1}^4\bigg[ \sum_{k=1}^4\bigg(  m_{lk} \sum_{j=1}^4 T_{kj}\rho_{jl} \bigg) \bigg] \bigg)^2
\end{equation}
Eq. \ref{quadratic} forms a second order polynomial in terms of the input elements. It is easy to show \cite{robustness} that the entanglement of formation for a two qubit pure state can be rewritten as a quadratic also:
\begin{equation}
E(|\psi\rangle) = 4a^2d^2 + 4b^2c^2 - 8abcd\cos(\theta_3 - \theta_2 - \theta_1)
\end{equation}
\noindent where the (real) numbers $\{ a,b,c,d, \theta_1, \theta_2, \theta_3 \}$ are defined in Eq. \ref{eq:purestate}. So it should not be altogether surprising that the QNN can learn an entanglement indicator for the two-qubit pure system. But clearly this function, according to Theorems 1 and 2, ought to be able to be approximated by both the RVNN and the CVNN, especially for the relatively simple case where all the phase offsets, \{$\theta_{i}$\}, are zero.  As we will see, however, even this turns out to be no easy task for either classical net to learn.

%DETAILS OF THE DERIVATION - NOT SURE WE NEED TO INCLUDE THIS

%If one recall that the entanglement of formation for a pure state $| \psi \rangle = \sum_i a_i|e_i \rangle  $
%can be define as \cite{wootters 97} \cite{coffman}
%\begin{equation*}
%E( \psi ) = \epsilon\big[ C(\psi) \big ]
%\end{equation*}
%where $C(\psi) = \bigg| \sum_i a_i^2 \bigg| $ and
%\begin{equation*}
%\epsilon(x) = H( \frac{1}{2} + \frac{1}{2}\sqrt{1-x^2} \ \ \ \ 0 \leq x \leq 1
%\end{equation*}
%with the function $H$ being define as
%\begin{equation*}
%H(x) = -\big[ x\log_2(x) + (1-x)\log_2(1-x) \big]
%\end{equation*}\\
%
%
%For a two qubit system, a pure state $\psi$ can be written explicitly as
% $|\psi \rangle = \big| a, be^{i\theta_1} , ce^{i\theta_2}, de^{i\theta_3} \rangle $
%then the equations above will give us the entanglement of formation for a pure state two qubit system as
%\begin{equation*}
%E(\psi) = 4a^2d^2 + 4b^2c^2 - 8abcd\cos(\theta_3 - \theta_2 - \theta_1)
%\end{equation*}
%which is a continuous function. Therefore, we know that there exists a real-valaued neural network that can solve the entanglement of formation problem for a two qubit system. However, we don't know how many hidden neurons are needed and how many epoch must be done for the network to figure this out. \\

    \subsection{Results}

% The QNN method was able to learn the entanglement for two-qubit states and generalized to 5-qubit states with a minimal set of training data. It was also found that a classical ANN was not able to generalize the entanglement solution using the same minimal set of training data\cite{multi qubit}.
%
%The motivation behind using a CVNN for learning entanglement stems from the similarities between it and the well performing QNN, where the QNN system is inherently complex-valued. The CVNN also exhibits interference due to phase differences in inputs and weights, a key characteristic of quantum systems. Considering the similarities between the function of a CVNN and the QNN, it is desirable to benchmark the abilities of a CVNN relative to the RVNN and QNN at learning a measure for quantum entanglement.

In our previous work \cite{2008} we saw that the QNN was able to learn entanglement for a two-qubit system from a training set of only four pure states. It should also be noted that the QNN had also successfully generalized to mixed states, as well. Here we train, and test, only on pure states; nevertheless, neither the RVNN nor the CVNN was able to accomplish this. 

Training for each of the four nets on the entanglement witness is shown in Table \ref{table:entangletrain}. Each network was given the entire $4 \times 4$ density matrix of the state as input. The classical nets again were given a hidden layer, and a single output.  The architectures used for the NeuralWorks, RVNN, and CVNN were 16 (input layer), 8 (hidden layer), 1 (output layer). Numbers of epochs were 5 to 10 thousand for the real valued nets; 1000 for the CVNN, and only 20 for the QNN. Again we note the increase in efficiency of training in going from RVNN to CVNN, and in going from CVNN to QNN. For the minimal training set of four, all three classical nets trained below $1\%$ error. 

However the testing error, shown in Table \ref{table:entangletest} was quite bad for all the classical nets. Increasing the size of the training set did enable marginally better fitting by all the classical networks, but testing error remained an order of magnitude larger than with the QNN. Increasing $N_h$ did not help. Increasing the size of the training set affected training and testing very little for the QNN, which had already successfully generalized with a set of only four.  Neither the NeuralWorks nor the RVNN could learn pure state entanglement well, even with a training set as large as 100. And while the (``quantum-inspired'') CVNN could use interference, it did not generalize efficiently or well, either. This is not surprising when one considers that it is precisely access to the entangled part of the Hilbert space that the CVNN lacks; that is, the state of CVNN cannot itself be entangled. In that sense it is inherently classical.

    \def\arraystretch{1.5}

    \begin{table}[h]
    \centering
    \begin{tabular}{l|l|l|l|l|}
    \cline{2-5}
    \multicolumn{1}{c|}{}  & \multicolumn{4}{c|}{Training RMS error (\%)}    \\ \hline
    \multicolumn{1}{|c|}{Training Pairs} & \multicolumn{1}{c|}{N.Works} & \multicolumn{1}{c|}{RVNN} & \multicolumn{1}{c|}{CVNN} & \multicolumn{1}{c|}{QNN}  \\ \hline
    \multicolumn{1}{|l|}{100} & 5.66 & 3.74 & 0.97 & 0.04   \\ \hline
    \multicolumn{1}{|l|}{50} & 5.96 & 5.89 & 0.53 & 0.09  \\ \hline
    \multicolumn{1}{|l|}{20} & 6.49 & 4.97 & 0.04 & 0.2  \\ \hline
    \multicolumn{1}{|l|}{4}  & 0.00 & 0.93 & 0.01 & 0.2  \\ \hline
    \end{tabular}
    \caption{Training on entanglement of pure states with zero offsets, for the NeuralWorks, RVNN, CVNN, and QNN.}
    \label{table:entangletrain}
    \end{table}

    \begin{table}[h]
    \centering
    \begin{tabular}{l|l|l|l|l|}
    \cline{2-5}
    \multicolumn{1}{c|}{}  & \multicolumn{4}{c|}{Testing RMS error(\%)}    \\ \hline
    \multicolumn{1}{|c|}{Training Pairs} & \multicolumn{1}{c|}{N.Works} & \multicolumn{1}{c|}{ANN} & \multicolumn{1}{c|}{CVNN} & \multicolumn{1}{c|}{QNN}  \\ \hline
    \multicolumn{1}{|l|}{100} & 7.56 & 5.39 & 3.61 & 0.2   \\ \hline
    \multicolumn{1}{|l|}{50} & 7.91 & 10.7 & 6.00 & 0.3  \\ \hline
    \multicolumn{1}{|l|}{20} & 13.6 & 15.5 & 9.48 & 0.4  \\ \hline
    \multicolumn{1}{|l|}{4}  & 48.2 & 51.9 & 55.0 & 0.4  \\ \hline
    \end{tabular}
    \caption{Testing on entanglement of pure states. Each network was tested on the same set of 25 randomly chosen pure states with zero offset. }
        \label{table:entangletest}
    \end{table}

\section{Conclusions and Future Work}

The problem of quantifying speedup for quantum algorithms continues to be a difficult one, solved only on a case-by-case basis \cite{shor,grover,bernstein}. The situation with quantum machine learning is similar \cite{Biamonte,Aimeur}. We need both general theorems \cite{namthesis} and benchmarking for specific architectures and learning methods.

The marriage of quantum computing and machine learning can be enormously beneficial to both subfields: with machine learning techniques we can find ``algorithms'' \cite{2008}, do automatic scale-up \cite{multi qubit}, and increase robustness significantly \cite{2 qubit noise,robustness}; while with quantum systems and their access to the enormously greater Hilbert space, machine learning can be both more efficient and more powerful \cite{2008}. In this paper we provide evidence of both. We have shown, here, that a fully quantum neural network can do standard classical benchmarking tasks more efficiently than either a real-valued or a complex-valued classical net. Moreover, a fully QNN can do {\it quantum} calculations efficiently, that is, without explicit simulation. In any physical implementation there will also be additional, natural, nonlinearities we can take advantage of, and, possibly, control and use for machine learning.

There is currently much work being done using ``quantum-inspired'' machine learning \cite{bhattacharyya}. Certainly it is of great value to tease out exactly whence the efficiency, and the complexity, of quantum computing - and specifically quantum neural computing - arise. One way of doing this is precisely to do ``quantum-inspired'' calculations: to allow, e.g., the state of the system to be complex-valued in some sense. Clearly without allowing full superposition this will not be fully quantum mechanical, because it will not include entangled states. There are many more states that are at least partially entangled than there are classical states (in the sense of being product states of subsystems.) But fully quantum mechanical calculations can quickly become dauntingly enormous, so the question is, how much can we get from how little?

There are, of course, many more kinds of machine learning tasks that we did not explore here. Moreover, ours is by no means the only quantum machine learning architecture \cite{Biamonte, chen} or approach \cite{Amin} possible. Chen et al. \cite{chen} also found a quantm advantage for their (very different) quantum neural network model, but, as pointed out in the Introduction, this is not a universal finding \cite{Ronnow}. It will be of great interest to see what kinds of problems can benefit significantly from a fully quantum mechanical approach. It is not at all obvious what these are, or how well simulations will translate to physical implementation. There is much still to do.

%``let us sit down and wait together!'' (my hero)

% if have a single appendix:
%\appendix[Proof of the Zonklar Equations]
% or
%\appendix  % for no appendix heading
% do not use \section anymore after \appendix, only \section*
% is possibly needed

% use appendices with more than one appendix
% then use \section to start each appendix
% you must declare a \section before using any
% \subsection or using \label (\appendices by itself
% starts a section numbered zero.)
%

%\appendices
%\section{Proof of the First Zonklar Equation}
%Some text for the appendix.

% use section* for acknowledgement
\section*{Acknowledgment}

The authors would like to thank William Ingle for helpful discussions.

% Can use something like this to put references on a page
% by themselves when using endfloat and the captionsoff option.
\ifCLASSOPTIONcaptionsoff
  \newpage
\fi

% trigger a \newpage just before the given reference
% number - used to balance the columns on the last page
% adjust value as needed - may need to be readjusted if
% the document is modified later
%\IEEEtriggeratref{8}
% The "triggered" command can be changed if desired:
%\IEEEtriggercmd{\enlargethispage{-5in}}

% references section

% can use a bibliography generated by BibTeX as a .bbl file
% BibTeX documentation can be easily obtained at:
% http://www.ctan.org/tex-archive/biblio/bibtex/contrib/doc/
% The IEEEtran BibTeX style support page is at:
% http://www.michaelshell.org/tex/ieeetran/bibtex/
%\bibliographystyle{IEEEtran}
% argument is your BibTeX string definitions and bibliography database(s)
%\bibliography{IEEEabrv,../bib/paper}

\begin{thebibliography}{1}

\bibitem{feynman1} R.P. Feynman (1982), {\it Simulating physics with computers}, Int. J. of Theo. Phys. {\bf 21}, 467.

\bibitem{shor} P.W. Shor (1994), {\it Algorithms for quantum computation: discrete logarithms and factoring}, Proceedings of the 35th Annual Symposium on Foundations of Computer Science (IEEE)

\bibitem{grover} L.K. Grover (1996), {\it A fast quantum mechanical algorithm for data base search}, Proceedings of the 28th Annual ACM Symposium on the Theory of Computing, 212.

\bibitem{bernstein} E. Bernstein and U. Vazirani (1997) {\it Quantum complexity theory}, SIAM J. Comput. {\bf 26}, pp 1411-1473.

\bibitem{bravyi} S. Bravyi, D. Gosset, and R. Konig (2017), {\it Quantum advantage with shallow circuits}, arXiv:1704.00690v1.

\bibitem{Ronnow} T.F. Ronnow, Z. Wang, J. Job, S. Boixo, S.V. Isakov, D. Wecker, J.M. Martinis, D.A. Lidar, and M. Troyer (2014), {\it Defining and detencting quantum speedup}, Science {\bf 345}, 420.

\bibitem{existence} A.R. Calderbank and P. Shor (1996) {\it Good quantum error correction codes exist}, Phys. Rev. A {\bf 54}, pp 1098-1105.

\bibitem{existence2} G. Ortiz, J. Gubernatis, E. Knill, and R. Laflamme (2001), {\it Quantum algorithms for femionic simulations} Phys. Rev. A {\bf 64} 022319.

\bibitem{orig1}  R. Chrisley (1995), {\it Quantum Learning}, In New directions in cognitive science: Proceedings of the international symposium, Saariselka, Finland, P. Pylkkänen and P. Pylkkö (editors). Finnish Association of Artificial Intelligence, Helsinki, 77-89.  

\bibitem{orig2} S. Kak, {\it On quantum neural computing}, Advances in Imaging and Electron Physics {\bf 94}, 259 (1995).

\bibitem{orig3} E.C. Behrman, J. Niemel, J.E. Steck, and S.R. Skinner (1996), {\it A quantum dot neural network}, Proceedings of the Fourth Workshop on Physics and Computation (PhysComp96), 22.

\bibitem{schutzhold} R. Schutzhold (2002), {\it Pattern recognition on a quantum computer}, arXiv:0208063v3.

\bibitem{trugenberger} C.A. Trugenberger (2002), {\it Pattern recognition in a quantum computer}, arXiv: 0210176v2 .

\bibitem{schuld} M. Schuld, I. Sinayskiy and F. Petruccione (2016), {\it Prediction by linear regression on a quantum computer}, Phys. Rev. A {\bf 94}, 022342.

%\bibitem{wang} G. Wang (2017) {\it Quantum algorithm for linear regression}, Phys. Rev. A {\96}, 012335.  

\bibitem{schuldrev} M. Schuld, I. Sinayskiy and F. Petruccione (2015), {\it An introduction to quantum machine learning} Contemp. Phys. {\bf 56}, 172.

\bibitem{arunaschalam} S. Arunaschalam and R deWolf (2017), {\it A survey of quantum learning theory}, arXiv:1701.06806v3.

\bibitem{Biamonte} J. Biamonte, P. Wittek, N. Pancotti, P. Rebentrost, N. Wiebe, and S. Lloyd (2018), {\it Quantum machine learning}, arXiv:1611.09347v2.

\bibitem{Aimeur} E. Aimeur, G. Brassard, and S. Gambs (2013), {\it Quantum speed-up for unsupervised learning}, Mach. Learn {\bf 90}, 261-87.

\bibitem{2002} E.C. Behrman, V Chandrashekar, Z. Wang, C.K. Belur, J.E. Steck, and S.R. Skinner (2002), {\it A quantum neural network computes entanglement}, arXiv:0202131.

\bibitem{2008} E.C. Behrman, J.E. Steck, P. Kumar, and K.A. Walsh (2008), {\it Quantum algorithm design using dynamic learning}, Quantum Inf. Comput. {\bf 8} pp. 12-29

\bibitem{Hentschel} A. Hentschel and B.C Sanders (2010), {\it Machine learning for precise quantum measurement}, arXiv:0910.0762v2.

\bibitem{chen} J. Chen, L. Wang, and E. Charbon (2017), {\it A quantum-omplementable neural network model}, Quantum Inf. Process. {\bf 16}, 245.

\bibitem{multi qubit} E. C. Behrman and J. E. Steck (2013), {\it Multiqubit entanglement of a general input state}. Quantum Inf. Comput. {\bf 13}, 1-2, pp. 36-53.

\bibitem{wiebe} N. Wiebe, C. Granade, and D.G. Cory (2015), {\it Quantum bootstrapping via compressed quantum Hamiltonian learning}, arXiv: 1409.1524v3.

\bibitem{Cai} X.D. Cai, D. Wu, Z.E. Su, M.C. Chen, X.L. Wang, L. Li, N.L. Liu, C.Y. Lu, and J.W. Pan (2015), {\it Entanglement-based machine learning on a quantum computer}, Phys. Rev. Lett. {\bf 114}, 110504.

\bibitem{upcoming} N.L. Thompson, N.H. Nguyen, E.C. Behrman, and J.E. Steck (2018), {\it Reverse engineering pairwise entanglement}, Quant. Inf. Comput. (to be submitted).

\bibitem{neilsen}  M. A. Nielsen and I. L. Chuang (2001), {\it Quantum Computation and Quantum Information}, Cambridge: Cambridge University Press.

% E.C. Behrman, R.E.F. Bonde, J.E. Steck, and J.F. Behrman (2014), {\it On the correction of anomalous phase oscillation in entanglement witnesses using quantum neural networks}, IEEE Trans. on Neural Networks and Learning Systems {\bf 25}, pp 1696-1703.

\bibitem{2 qubit noise} E.C. Behrman, N.H. Nguyen, J.E. Steck, M. McCann (2016), {\it Quantum neural computation of entanglement is robust to noise and decoherence}, in Quantum Inspired Computational Intelligence: Research and Applications, S. Bhattacharyya, ed. (Morgan Kauffman, Elsevier) pp. 3-33.

\bibitem{robustness} N.H. Nguyen, E.C. Behrman, and J.E. Steck (2016), {\it Robustness of quantum neural calculation increases with system size}, arXiv:1612.07593

\bibitem{cybenko} G. Cybenko (1989) { \it Approximation by superpositions of a sigmoidal function}, Math. Control Signal Systems {\bf 2}, pp.  303-314.

\bibitem{NeuralWorks} Neuralware (2000), {\it Getting started: a Tutorial in NeuralWorks Professional II/Plus}

\bibitem{cnn} Taehwan Kim and Tülay Adali (2001), {\it Complex backpropagation neural network using elementary transcendental activation function} Proc. of IEEE Int. Conf. on Acoustics, Speech, and Signal Processing, 7-11 May 2001 {\bf 2} pp. 1281-1284; Taehwan Kim and Tülay Adali (2003) {\it Approximation by fully complex multilayer perceptrons}, Neural Comput. {\bf 15} pp. 1641-1666. DOI=http://dx.doi.org/10.1162/089976603321891846

\bibitem{complex network text} I. Aizenberg (2011), {\it Complex-Valued Neural Networks with Multi-Valued Neurons}, Berlin-Heidelberg: Springer.

\bibitem{nams frat} N.H. Nguyen, E.C. Behrman and J.E. Steck, {A universal physically implementable quantum neural network}, in preparation.

\bibitem{peres} A. Peres (1995), {\it Quantum theory: concepts and methods}. Dordrecht: Kluwer Academic Press.

\bibitem{werbos}  Paul Werbos, in {\it Handbook of Intelligent Control}, Van Nostrand Reinhold, p. 79 (1992);  Yann le Cun, {\it A theoretical framework for back-propagation} in {\it Proc. 1998 Connectionist Models Summer School,} D. Touretzky, G. Hinton, and T. Sejnowski, eds., Morgan Kaufmann, (San Mateo), pp. 21-28 (1988).

\bibitem{namthesis} N.H. Nguyen (2019), {\it Complexity and power of quantum neural networks}, Ph.D. Thesis, Wichita State University (in progress.)

\bibitem{bhattacharyya} S. Dey, S. Bhattacharyya, and U. Maulik (2014), {\it Quantum inspired genetic algorithm and particle swarm optimization using chaotic map model based interference for gray level image thresholding}, Swarm and Evol. Comput. {\bf 15} pp 38-57; S. Dey, S. Bhattacharyya, and U. Maulik (2017), {\it Efficient quantum inspired meta-heuristics for multi-level true colour image thresholding}, Appl. Soft Comput. {\bf 56} pp 472-513; and references cited therein.

\bibitem{irisdataset} Available at $https://en.wikipedia.org/wiki/Iris_flower_data_set$

\bibitem{perceptron} K.-L. Seow, E.C. Behrman, and J.E. Steck (2015), {\it Efficient learning algorithm for quantum perceptron unitary weights}, arXiv:1512.00522

\bibitem{harris} D. Harris and S. Harris (2012) {\it Digital design and computer architecture (2nd ed.)}, San Francisco, Calif.: Morgan Kaufmann, p. 129.

\bibitem{fisher} R. A. Fisher (1936), {\it The use of multiple measurements in taxonomic problems} Ann. Eugenics {\bf 7}, pp 179–188.

\bibitem{ecb1983} E.C. Behrman, G.A. Jongeward, and P.G. Wolynes (1983), {\it A Monte Carlo approach for the real time dynamics of tunneling systems in condensed phases}
J. Chem. Phys. {\bf 79}, 6277-6281.

\bibitem{feynman} R.P. Feynman, R.B. Leighton, and M. Sands (1964), {\it The Feynman Lectures on Physics}, Addison Wesley, Reading, MA, Vol III. 
%Also available online at: http://www.feynmanlectures.caltech.edu/III_toc.html

\bibitem{optimizations} V. Vedral, M.B. Plenio, M.A. Rippin, and P.L. Knight (1997) {\it Quantifying entanglement} Phys. Rev. Lett. {\bf 78}, pp. 2275-2279 (1997);  S. Tamaryan, A. Sudbery, and L. Tamaryan (2010), {\it Duality and the geometric measure of entanglement of general multiqubit W states} Phys. Rev. A {\bf 81}, 052319

\bibitem{wootters} W.K. Wootters (1998), { \it Entanglement of Formation of an Arbitrary State of Two Qubits } Phys. Rev. Lett. {\bf 80}, 2245.

\bibitem{Amin} M.H. Amin, E. Andriyash, J. Rolfe, B. Kulchytskyy, and R. Melko (2018), {\it Quantum Boltzmann machine}, Phys. Rev. X {\bf 8}, 021050.

%
%\bibitem{coffman} V. Coffman, J. Kundu, and W.K. Wootters (2000), {\it Distributed entanglement}, Phys. Rev. A {\bf 61}, 052306. \\










\end{thebibliography}
%
% <OR> manually copy in the resultant .bbl file
% set second argument of \begin to the number of references
% (used to reserve space for the reference number labels box)

% biography section
%
% If you have an EPS/PDF photo (graphicx package needed) extra braces are
% needed around the contents of the optional argument to biography to prevent
% the LaTeX parser from getting confused when it sees the complicated
% \includegraphics command within an optional argument. (You could create
% your own custom macro containing the \includegraphics command to make things
% simpler here.)
%\begin{biography}[{\includegraphics[width=1in,height=1.25in,clip,keepaspectratio]{mshell}}]{Michael Shell}
% or if you just want to reserve a space for a photo:

\begin{IEEEbiography}[{\includegraphics[width=1in,height=1.25in,clip,keepaspectratio]{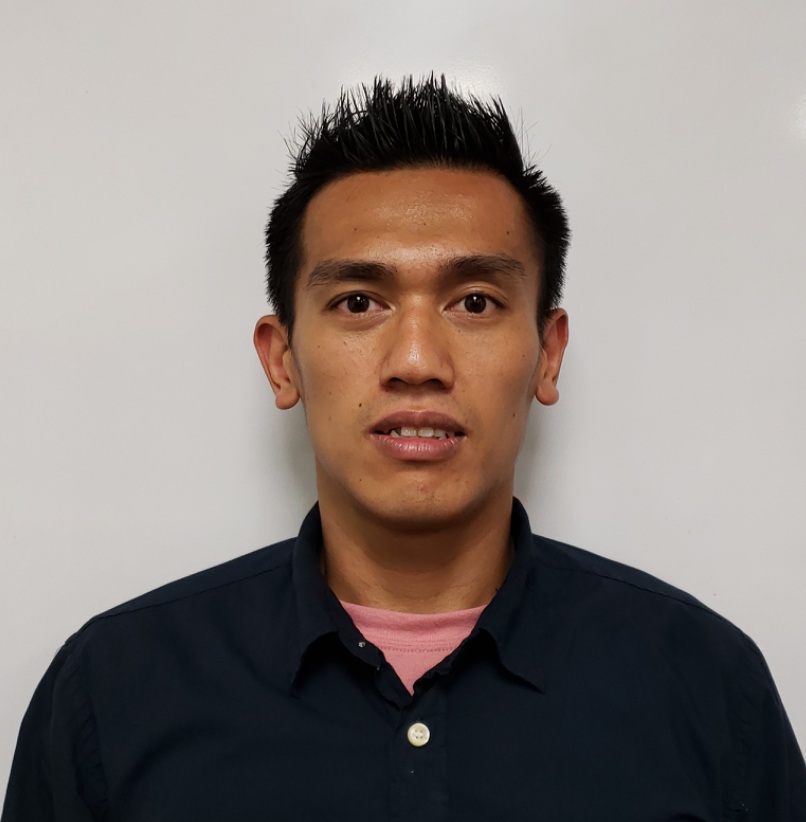}}]{Nam H. Nguyen}
earned a B.S in Mathematics from Eastern Oregon University, and an M.Sc in Applied Mathematics from Wichita State University. He is currently a Ph.D candidate in applied mathematics at Wichita State University under the supervision of professor E.C. Behrman. His research interests and publications are in the area of quantum neural networks and machine learning, partial differential equations, and graph theory. He has done research work in inorganic and physical chemistry as well.

\end{IEEEbiography}

\begin{IEEEbiography}[{\includegraphics[width=1in,height=1.25in,clip,keepaspectratio]{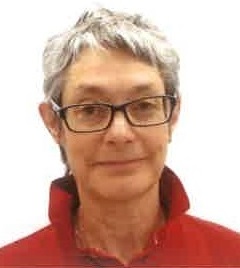}}]{E.C. Behrman}
%\begin{IEEEbiographynophoto}{E.C. Behrman}
earned an Sc.B from Brown University in mathematics, and an M.S. in chemistry and Ph.D. in physics from University of Illinois at Urbana-Champaign.  She did postdoctoral work at SUNY Stony Brook, and worked for four years at NYS College of Ceramics at Alfred University. She is currently professor of mathematics and of physics at Wichita State University. Her research interests and publications are broad, with over 80 papers in subjects ranging from chemical kinetics and reaction pathways to ceramic superconductors to nuclear waste vitrification. She was the first to predict the stability of inorganic buckyballs and buckytubes, and among the first to design and computationally test models for quantum neural networks and machine learning.
\end{IEEEbiography}

\begin{IEEEbiography}[{\includegraphics[width=1in,height=1.25in,clip,keepaspectratio]{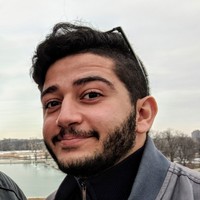}}]{Mohamed A. Moustafa}

holds a B.S. in Aerospace Engineering from Wichita State
University and B.S. in Physics from Emporia State University. His research interests
include machine learning, quantum computations, aviation and space applications; he
particularly enjoys bridging the gap between academia and software development. He
has contributed to research on aircraft loss-of-control prediction and developed
pilot-assistive warning displays using augmented reality devices at Wichita State
University. He has experience implementing various machine learning models including
GA for solving the facility layout problem and various implementations of artificial
neural networks. Mohamed is currently an Applications and Front-End engineer at a
cloud HPC (High Power Computations) company based in San Francisco, CA.

\end{IEEEbiography}

\begin{IEEEbiography}[{\includegraphics[width=1in,height=1.25in,clip,keepaspectratio]{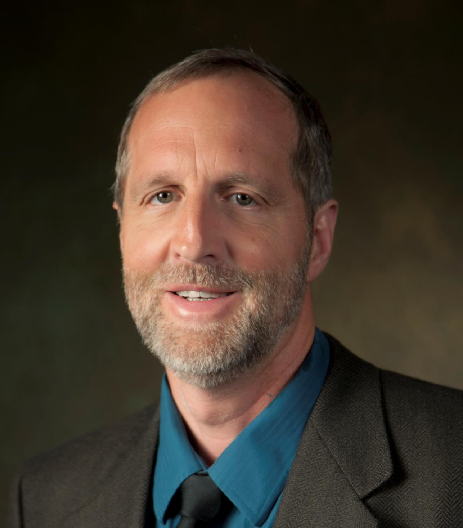}}]{J.E. Steck}
%\begin{IEEEbiographynophoto}{J.E. Steck}
has a B.S. and M.S. in Aerospace Engineering and a Ph.D. in Mechanical Engineering from the University of Missouri-Rolla where he performed research in the use of finite element methods in mechanics, fluids and aero-acoustics. He has done postdoctoral work in artificial neural networks as control systems for the Intelligent Systems Center at the University of Missouri-Rolla, and worked for two years at McDonnell Aircraft Company in the Aerodynamics department doing flight dynamics support for flight simulation, wind tunnel and flight testing of the AV-8B aircraft.  He is currently professor of Aerospace Engineering at Wichita State University where he has taught for twenty-three years.  He teaches undergraduate and graduate courses in flight dynamics and controls, artificial neural networks, and computational methods.   His current work includes: intelligent adaptive control systems for general aviation aircraft, integrated aircraft structural health monitoring, quantum neural computing, optical aircraft ice detection, optical neural computing, and the use of artificial neural networks for system modeling and control.
\end{IEEEbiography}

% You can push biographies down or up by placing
% a \vfill before or after them. The appropriate
% use of \vfill depends on what kind of text is
% on the last page and whether or not the columns
% are being equalized.

%\vfill

% Can be used to pull up biographies so that the bottom of the last one
% is flush with the other column.
%\enlargethispage{-5in}

% that's all folks
\end{document}